\documentclass[twocolumn]{aastex62}
\usepackage{amsmath,amssymb,amsfonts}
\usepackage{graphicx}
\usepackage{hyperref}
\usepackage{physics}

\newcommand{\inR}{\in \mathbb{R}}

\begin{document}

\title{Roles of crust and core in the tidal deformability of neutron stars}

\author{A. M. Kalaitzis}

\author{T. F. Motta}
\author{A. W. Thomas}
\affiliation{CSSM and ARC Centre of Excellence for Particle Physics at the Terascale, Department of Physics, University of Adelaide SA 5005 Australia}
\submitjournal{ApJL}
\begin{abstract}
With the recent measurement of GW170817  providing constraints on the tidal deformability of a neutron star, it is very important to understand what features of the equation of state have the biggest effect on it. We therefore study the contribution of the crust to the tidal deformability and the moment of inertia of a neutron star for a variety of  well-known equations of state. It is found that the contributions to these quantities from the low density crust are typically quite small and as a result the determination of the tidal deformability provides an important constraint on the equation of state of dense matter.
\end{abstract}

\keywords{stars: neutron, gravitational waves, equation of state}

\section{Introduction}

Neutron stars (NS) are of enormous interest as their cores are the repositories of the densest nuclear matter in the Universe. Exploring the nature of nuclear matter at densities many times that of nuclear matter is one of the most fascinating challenges in strong interaction physics, not the least because of the uncertainties surrounding the appearance of hyperons, $\Delta$'s and even quark matter. Of course, these dense cores are surrounded by a crust which contains low density nuclear matter which is expected to exhibit fascinating phase changes~\cite{WEBER2005193,10.1093/mnras/stx1510,glendenning_2000}. The low density region (LDR), corresponding to the crust, poses challenges with regards to modelling because of these complex phase structures as well as the variety of finite nuclei far from stability that may appear there~\cite{Chamel2008}.

Following the recent GW170817 measurement \cite{Abbott:2018wiz,Abbott:2018exr}, there has been enormous interest in studying the tidal deformability (TD) and the moment of inertia (MoI) of NS \cite{Wei:2018dyy,De:2018uhw,Han:2018mtj,Carson:2018xri,Zhao:2018nyf,Qian:2018qhh,Landry:2018jyg}, across a variety of equations  of state (EoS). Indeed, the TD is the key parameter extracted from the gravity wave observation of the NS merger and the bound produced in Refs.~\cite{Abbott:2018wiz,Abbott:2018exr,De}, namely $70\leq\Lambda_{1.4}\leq 580$, has been analysed recently~\cite{Landry:2018jyg} via the I-Love-Q relations to produce a constraint on the MoI of $8.82\leq\bar{I}\leq 14.74$ for a NS of mass $M=1.338M_\odot$ (where $\bar{I} = I/M^3$).

In this paper, we explore the proportional contribution that the LDR of the EoS makes to the TD and MoI for a selection of commonly used EoS which yield stars with a wide variety of radii and masses. Our conclusion is that the LDR corresponding to the crust makes a surprisingly small contribution to both the TD and the MoI and hence the constraints imposed by the analysis of the gravitational wave observations are of direct relevance to the determination of the EoS of dense matter.

\section{Theoretical Framework}

\begin{figure*}
    \centering
    \includegraphics[width=0.75\textwidth]{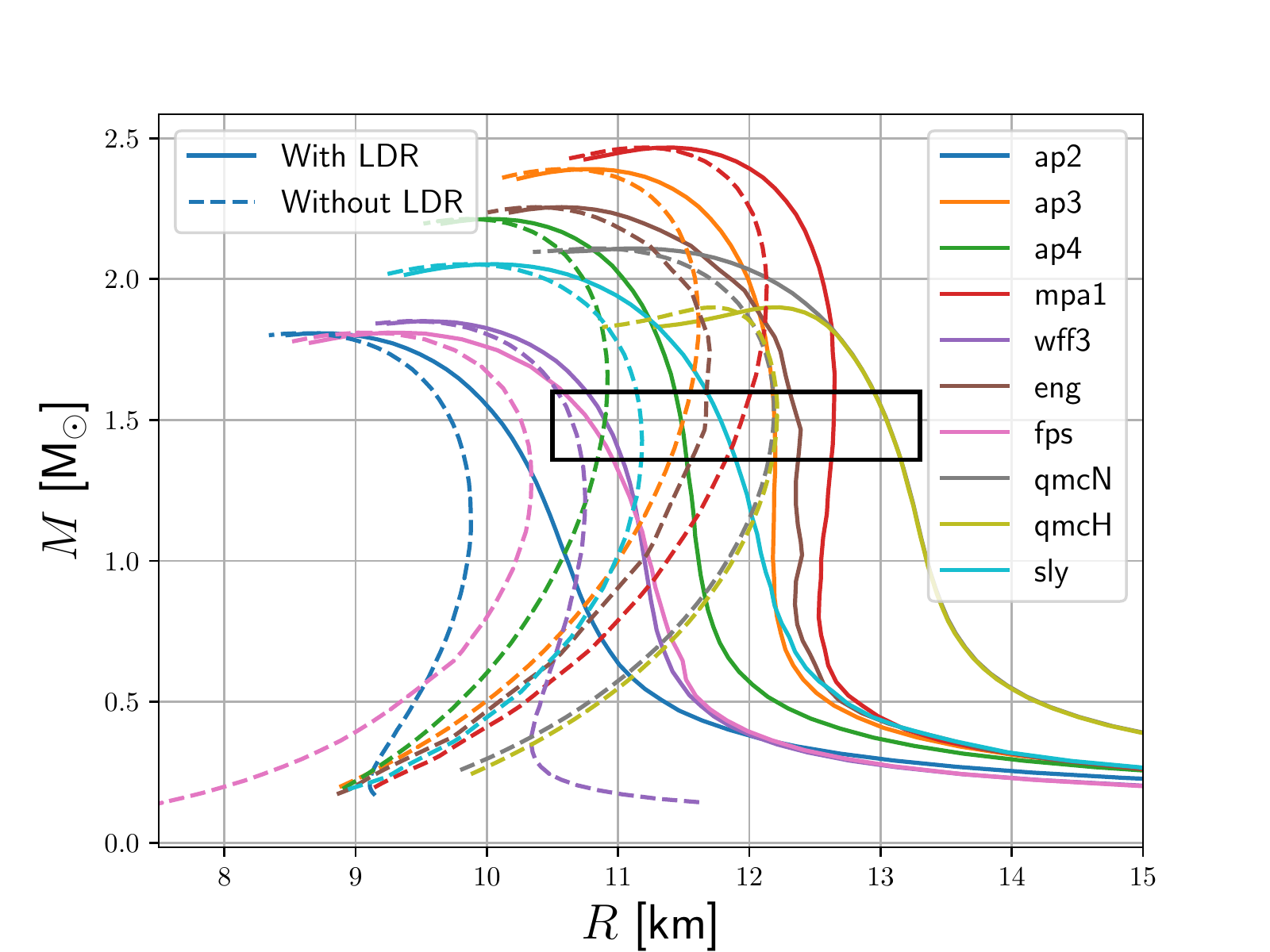} 
    \caption{Mass radius relation for the broad range of EoS studied here. 
The box represents the GW170817 90\% confidence constrained mass 
region~\cite{GBM,Abbott:2018wiz,Abbott:2018exr}.}
    \label{fig:MvsR}
\end{figure*}

We study eight EoS obtained from the online data base in Ref.~\cite{neutronstars}, which details all the models in question, as well as using two EoS calculated within the Quark Meson Coupling (QMC) model, discussed in Refs.~\cite{RIKOVSKASTONE2007341,GUICHON1996349,Motta:2019tjc}. These EoS include examples with and without hyperons and lead to stars with maximum masses varying over the range 1.8 to $2.5 M_\odot$ and 
with radii between 9 and 14km, as shown in Fig.~\ref{fig:MvsR}. This broad spread of NS properties gives us confidence that the conclusions drawn from our study are essentially model independent.
To calculate the NS profile, ordinarily one would match the model EoS description of the core to some low density EoS which accounts for the NS crust~\cite{10.1093/mnras/stx1510,Chamel2008}. In the region of energy density above 100 MeV/fm$^{3}$, we have fit each EoS using a polytropic form
\begin{align}
  \label{eq:poly}
  P(\epsilon) = A\epsilon^B+C\epsilon^D,
\end{align}
where $A,B,C$ and $D\inR$. This form continued to zero density defines the EoS without LDR modelling, with the original EoS containing LDR modelling. The choice of 100 MeV/fm$^{3}$ as the boundary is typical of the values usually chosen and we have checked that using 50 MeV/fm$^{3}$ instead makes no significant difference to our conclusions. As we see in Fig.~\ref{fig:MvsR}, by comparing solid and dashed curves of the same colour, the inclusion or omission of the LDR region leads to dramatically different mass versus radius curves.

These fits are then used to analyse the proportional contributions to the TD through a direct comparison of their values. This is appropriate because the TD is defined as a boundary point and not something cumulative (e.g. an integral)~\cite{Postnikov:2010yn,1967ApJ...149..591T}, so the direct calculation of its LDR contribution is impossible. This is equivalent to modelling a core only NS, containing just degenerate nuclear matter. The LDR contribution for the MoI may be calculated directly as it is defined as an integral \cite{1967ApJ...150.1005H,1968ApJ...153..807H}.

The NS masses are obtained by numerically solving the Tolman-Oppenheimer-Volkoff (TOV) 
equations~\cite{PhysRev.55.374}
\begin{align}
  \dv{P}{r} &= -\frac{(P(r)+\epsilon(r))(m(r)+4\pi r^3P(r))}{r(r-2m(r))} \label{eq:dpdr},\\
  \dv{m}{r} &= 4\pi r^2\epsilon(r) \label{eq:dmdr},
\end{align}
until the point where the pressure vanishes, $P(r=R)=0$ ($R$ the radius of the NS). This procedure led to the  plots of mass versus radius shown in Fig.~\ref{fig:MvsR}, where we note the characteristic shrinking of the radius of the star at small mass in the case where it is taken to consist of degenerate nuclear matter alone.

We define the variation of a parameter $\alpha$ as $\Delta\alpha/\alpha$ (for non-cumulative quantities) in terms of the percentage difference of the parameter when compared to its LDR removed counterpart 
\begin{align}
  \label{eq:variation}
  \frac{\Delta\alpha}{\alpha} = \frac{\qty(\alpha_{\text{with LDR}}-\alpha_{\text{without LDR}})}{\alpha_{\text{with LDR}}}.
\end{align}
Note that in the case of the TD this requires a choice of the parameter which is fixed for the two cases. We choose to compare the TD for stars with the same mass, since that is the physical parameter which is usually best known. The MoI, $I$, is calculated using the methods described in Refs.~\cite{1967ApJ...150.1005H,1968ApJ...153..807H}.

\begin{figure*}[h!]
	\centering
	\begin{minipage}{0.49\textwidth}
		\centering
		\includegraphics[width=1.1\textwidth]{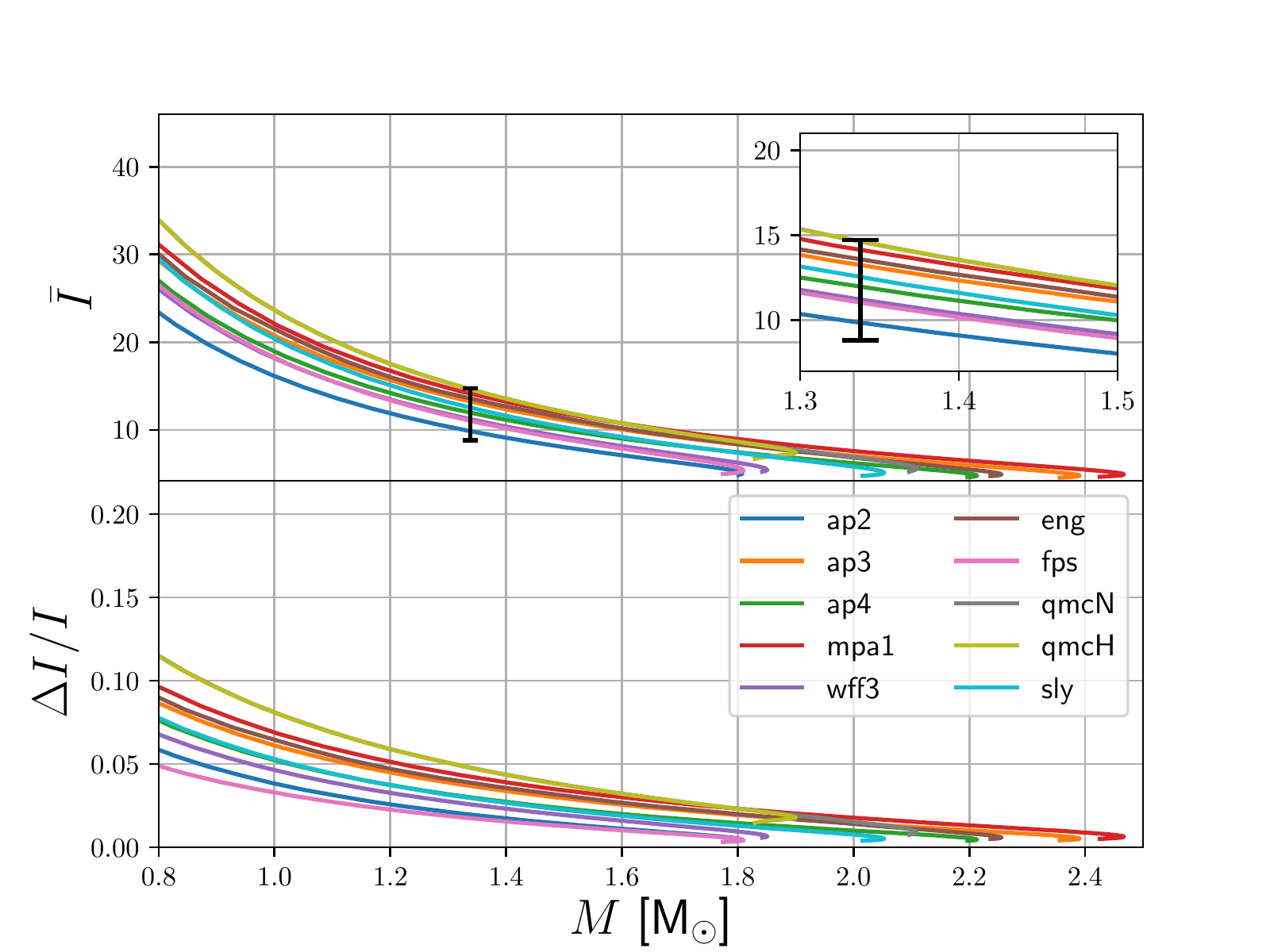} 
		\caption{The percentage of the MoI arising from the LDR is shown versus the NS total MoI. As explained in the text, this is calculated using the full EoS, including the LDR. The enlargement shows the constraint $8.82\leq\bar{I}\leq 14.74$ \cite{Landry:2018jyg}.}
		\label{fig:CrustMoI}
		\includegraphics[width=1.1\textwidth]{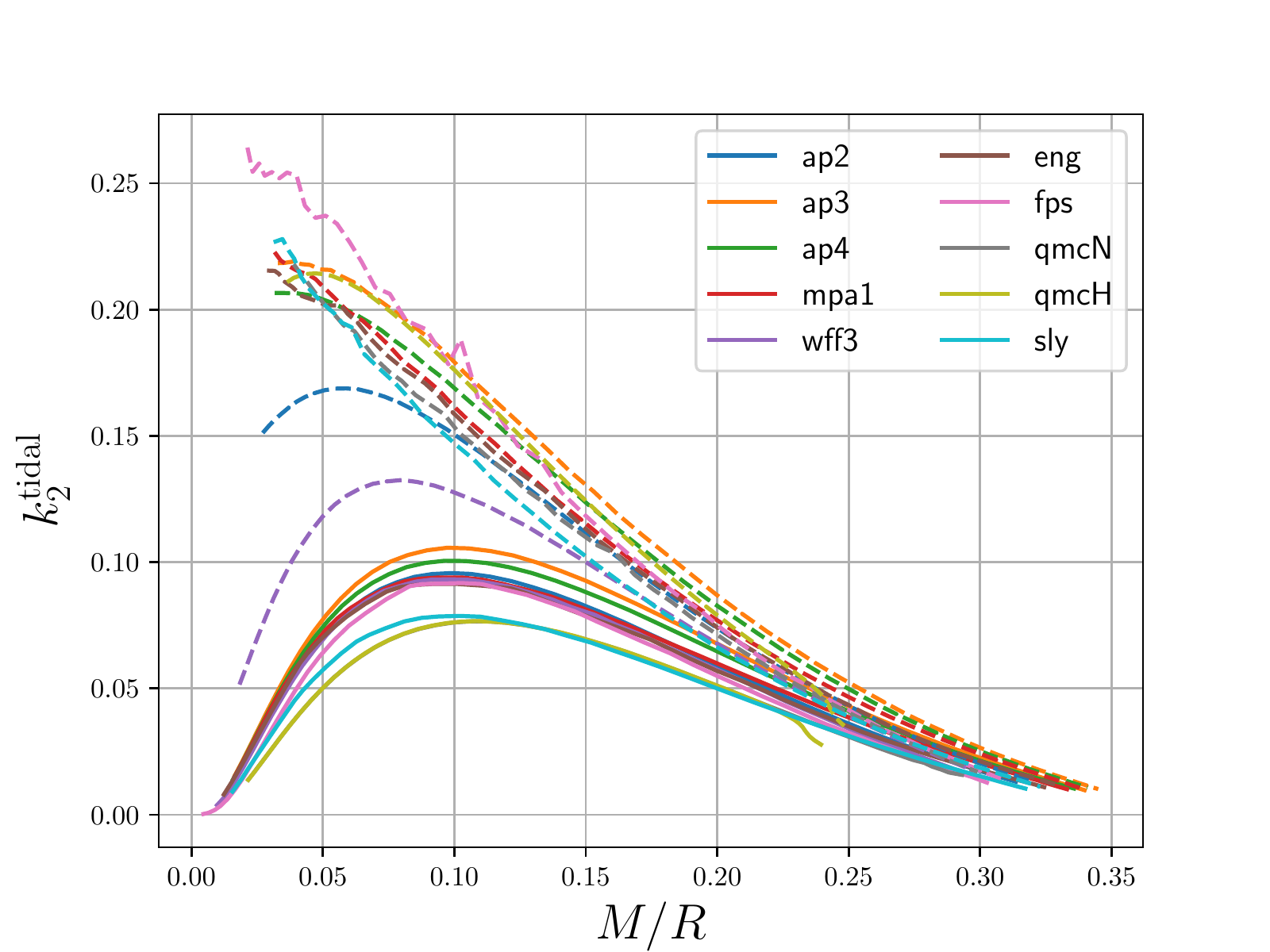}
		\caption{The tidal Love numbers $k_2^{\text{tidal}}$ are shown as a function of the compactness parameters of the stars, keeping same conventions as in Fig.~\ref{fig:MvsR}.}
		\label{fig:lovenumber}
	\end{minipage}\hfill
	\begin{minipage}{0.49\textwidth}
		\centering
		\includegraphics[width=1.1\textwidth]{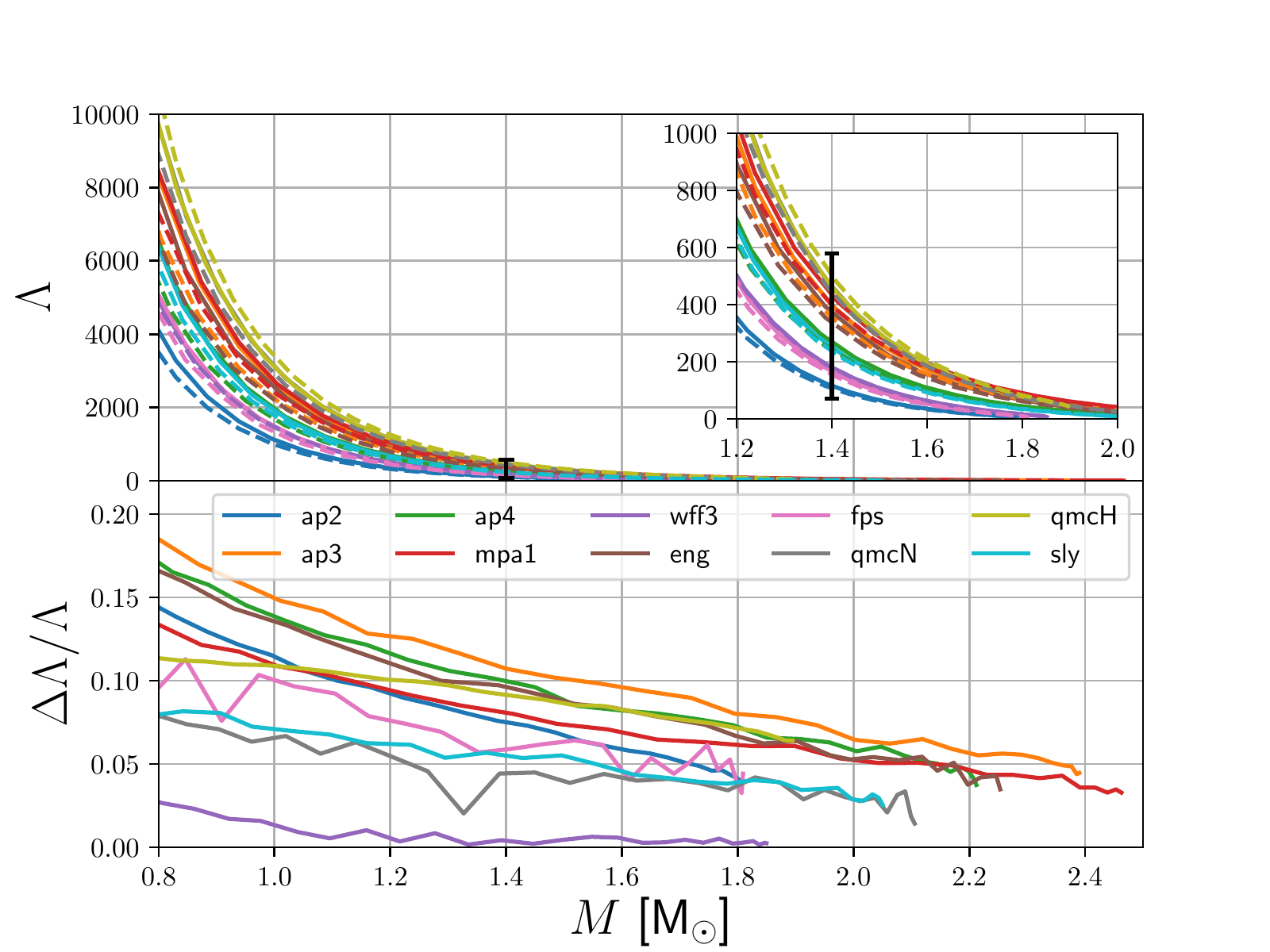} 
		\caption{Illustration of the percentage of the TD arising from the LDR versus the NS total TD for fixed NS mass. The enlargement shows a smaller range of masses, with the black bar representing the bound $70\leq\Lambda_{1.4}\leq 580$ \cite{Abbott:2018wiz,Abbott:2018exr}.}
		\label{fig:CrustTidal}
		\includegraphics[width=1.1\textwidth]{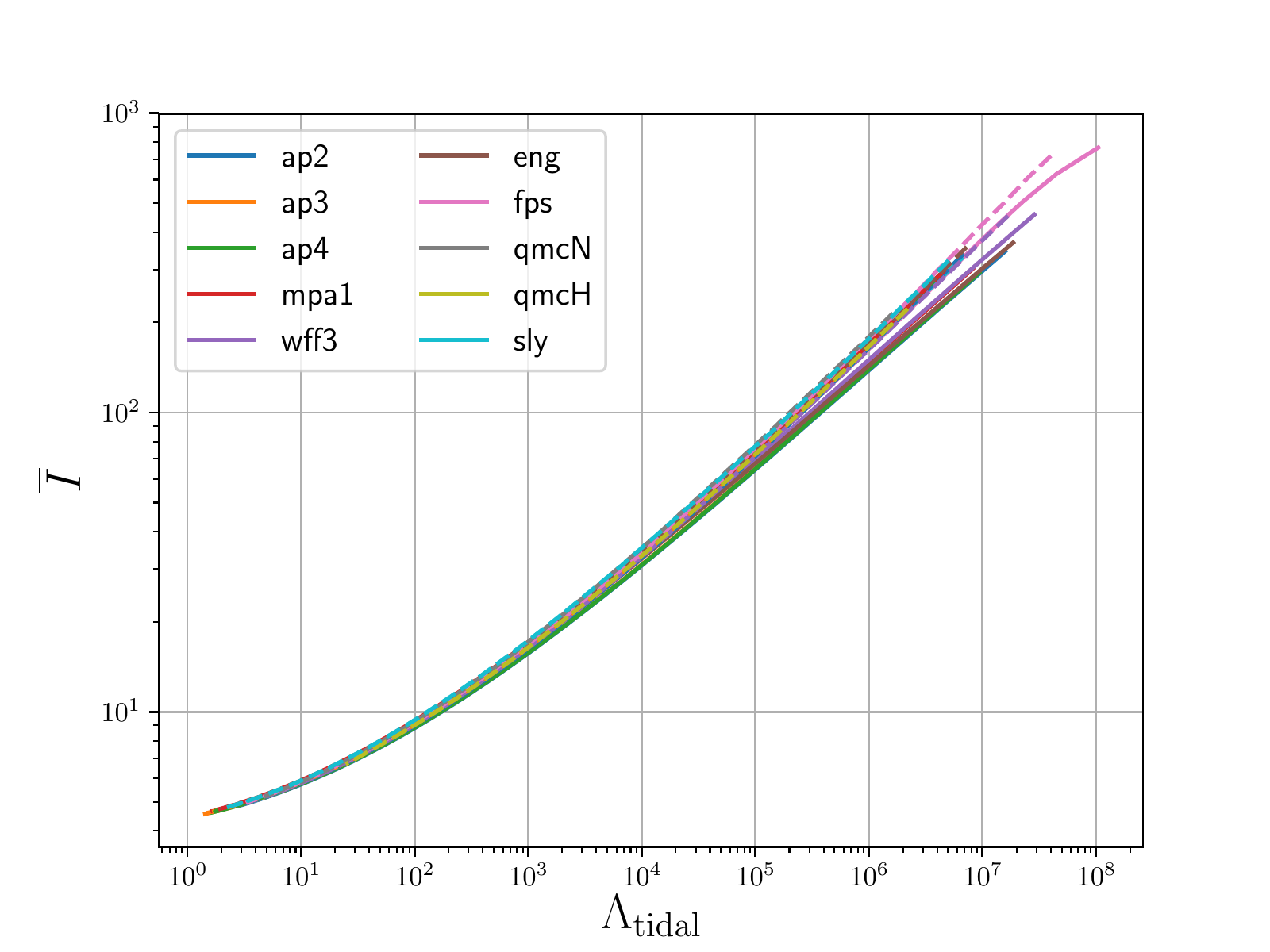}
		\caption{The I-Love relation is shown for all the EoS considered here, keeping same conventions as in Fig.~\ref{fig:MvsR}.}
		\label{fig:I-Love}
	\end{minipage}
\end{figure*}

The methods for calculating the tidal Love number, $k_2^{\text{tidal}}$, TD, $\Lambda$, and binary TD, $\widetilde{\Lambda}$, are those presented in Refs.~\cite{Postnikov:2010yn,1967ApJ...149..591T,Zhao:2018nyf} and summarised in the Supplementary Material of 
Ref.~\cite{Motta:2019tjc}. For future reference, the TD and binary TD are calculated using
\begin{align} \label{eq:td}
  \Lambda &= \frac{2}{3}k_2^{\text{tidal}}\beta^{-5},\\
  \widetilde{\Lambda} &= \frac{16}{13}\frac{(12q+1)\Lambda_1+(12+q)q^4\Lambda_2}{(1+q)^5},
\end{align}
where $M_2\leq M_1$ and $q=M_2/M_1$. Finally the I-Love relation is defined in Refs.~\cite{Yagi:2013awa,1967ApJ...150.1005H,1968ApJ...153..807H}.

\section{Results and Discussion}

Careful evaluation of the expressions in 
Refs.~\cite{Zhao:2018nyf,1967ApJ...150.1005H,1968ApJ...153..807H,Lattimer:2006xb} for the moment of inertia produces Fig.~\ref{fig:CrustMoI}, in which we show the dimensionless MoI and its crustal proportion, which is given by $\Delta I$ over $I$.

Note that for the range of masses of most practical interest~\cite{Lattimer:2006xb}, namely 
$M > 1.0 M_\odot$, the crustal contribution is found to be $\abs{\Delta I/I} < 9.0\%$. This value also decreases quickly as the mass increases. In particular, a star of solar mass $1.4M_\odot$ has $\abs{\Delta I/I} < 4.5\%$.

The contribution of the LDR to the tidal deformability is calculated by solving the equations presented in Refs.~\cite{Postnikov:2010yn,1967ApJ...149..591T} for every individual mass-radius curve shown in Fig.~\ref{fig:MvsR}. Using these results in Eq.~\ref{eq:variation} leads to the results presented in 
Fig.~\ref{fig:CrustTidal}.
It is found that $\abs{\Delta \Lambda/\Lambda} < 15\%$ for every NS with mass above $1.0 M_\odot$. For a typical NS of mass $1.4 M_\odot$ one concludes that the bounds are in fact
\begin{align}
  \label{eq:TDdiff}  \abs{\Delta I/I}_{M=1.4M_\odot} < 4.5\%&, \qquad \abs{\Delta \Lambda/\Lambda}_{M=1.4M_\odot} < 11\% \, .
\end{align}
We stress that our conclusion concerning the TD is independent of the visibly large difference in the mass radius relations (and consequentially the compactness parameter $\beta=M/R$) and tidal Love numbers shown in Fig.~\ref{fig:MvsR} and Fig.~\ref{fig:lovenumber}, respectively. The variance in the TD remains surprisingly low, in spite of the large changes in tidal Love numbers, because of the compensating effect of the large negative power from the $\beta$ term in Eq.~\ref{eq:td}.

The relatively small dependence of the TD on the presence of the crust is also reflected in a correspondingly small variation in the binary TD, because it depends linearly on $\Lambda_1$ and $\Lambda_2$ and involves only $M_1$ and $M_2$ as additional parameters.

Another interesting consequence of the removal of the low density EoS is its effects on the I-Love universal band. In Ref.~\cite{Yagi:2013awa} it was suggested that the universal I-Love relation could be a consequence of most EoS containing the same low density crust modelling and that the I-Love relation might possibly depend more on the low density region than the high.
However, as illustrated in Fig.~\ref{fig:I-Love}, the structure of the I-Love relation remains unchanged between EoS with and without the LDR for $M > 1.0M_\odot$. 

We stress that none of the results reported here should be interpreted as suggesting that the LDR is not crucial for a number of NS properties. As discussed in detail in Ref.~\cite{Piekarewicz:2018sgy} and illustrated here in Fig.~\ref{fig:MvsR}, the LDR contributes quite significantly to the radius for a star of a given mass. Furthermore, the imminent measurement by the NICER mission of the radius of a NS reaffirms that good modelling of the LDR will prove incredibly significant in obtaining reliable predictions for such an empirical result.

\section{Conclusion}

The present study of a wide variety of EoS suggests a surprising and important result. That is, while the crust of a NS does make a very important contribution to the radius of a star of given mass and hence to the compactness parameter, its contribution to the tidal deformability and moment of inertia is at the level of 10\% and 5\% or less, respectively, for stars of mass above $1.4 M_\odot$. This leads us to the conclusion that the measurement of binary tidal deformability in gravitational wave measurments does indeed provide a significant constraint on the EoS of dense nuclear matter.

\acknowledgments
This work was supported by the University of Adelaide and the Australian Research Council through grants DP150103101 and DP180100497.

\bibliographystyle{aasjournal}
\setcitestyle{authoryear,open={((},close={))}}

\end{document}